\documentstyle[a4,12pt,epsf]{article}
\newcommand{\noi}{\noindent}
\newcommand{\eq}{\begin{equation}}
\newcommand{\en}{\end{equation}}
\newcommand{\eqa}{\begin{eqnarray}}
\newcommand{\ena}{\end{eqnarray}}

\newcommand{\tr}{\mbox{Tr}\,}

\newcommand{\im}{\mbox{Im}\,}

\newcommand{\onabla}{{\overline \nabla}}

\newcommand{\bpartial}{{\bar \partial}}

\newcommand{\ttheta}{{\tilde \theta}}

\begin{document}

\renewcommand{\baselinestretch}{1.1}
\small\normalsize
\renewcommand{\theequation}{\arabic{section}.\arabic{equation}}
\renewcommand{\thesection}{\arabic{section}}
\language0

\hbox{}
\noindent February 2000 \hfill        

\vspace{0.5cm}
\begin{center}

\renewcommand{\thefootnote}{\fnsymbol{footnote}}
\setcounter{footnote}{0}

{\LARGE  Monopole classical solutions and the vacuum structure in
lattice gauge theories} 
\footnote{The work has been supported by the grant INTAS-96-370,
the RFRB grant 99-01-01230 and the JINR Dubna Heisenberg-Landau program.}

\vspace*{0.5cm}
{\large
V.K.~Mitrjushkin 
}\\

\vspace*{0.2cm}
{\normalsize
Joint Institute for Nuclear Research, 141980 Dubna, Russia
}     

\vspace{0.5cm}
Abstract.
\end{center}
Classical solutions corresponding to monopole--antimonopole pairs
are found in $3d$ and $4d$ $SU(2)$ and $U(1)$ lattice 
gauge theories.
The stability of these solutions in different theories is studied.

\section{Introduction}

One of the most interesting and still unresolved problems is the 
structure of a (nonperturbative) vacuum in QCD and a confinement 
mechanism. 

At the time being we have a number of different competing scenarios of
confinement.  One of the most promising approaches is a quasiclassical
approach promoted by Polyakov \cite{pol} which assumes that in the
treatment of infrared problems certain classical field configurations
are of paramount importance.  These classical field configurations
("pseudoparticles") are supposed to be stable, i.e.  they correspond to
local minima of the action and the interaction of these pseudoparticles
creates a correlation length which corresponds to a new scale --
confinement scale.  This approach gives a clear field--theoretical
prescription how to calculate analytically nonperturbative observables
in the weak coupling region.  In principle, this approach can be
extended to the case of "quasistable" solutions.

Another very attractive approach is a topological (or monopole) mechanism 
of confinement \cite{mon}. This mechanism suggests that the QCD vacuum 
state behaves like a magnetic (dual) superconductor, abelian
magnetic monopoles playing the role of Cooper pairs, at least, for
the specially chosen ("maximally abelian") gauge \cite{abe}.
At the time being this  approach remains the most popular one in 
numerical studies in lattice QCD. 

It is rather tempting to try to interprete lattice (abelian) monopoles
as pseudoparticles (stable or quasistable). 
Recently the classical solutions have been found which correspond to
Dirac sheet (i.e. flux tube) configurations \cite{ds}.
It is the aim of this note
to study the monopolelike ($M{\bar M}$) abelian solutions of classical
equations of motion in $~SU(2)~$ and $~U(1)~$ lattice gauge theories in
$d=3$ and $d=4$ dimensions.

In what follows periodic boundary conditions are presumed.
Lattice derivatives are~:
$~\partial_{\mu}f_x = f_{x+\mu}-f_x~$ ;
$~{\bar\partial}_{\mu}f_x = f_x-f_{x-\mu}~$
and $~\onabla_{\mu}(U) f_x = f_x 
-U^{\dagger}_{x-\mu;\mu}f_{x-\mu}U_{x-\mu;\mu}.$
$~~\Delta = -\sum_{\mu}\partial_{\mu}\bar\partial_{\mu}~$
and the lattice spacing is chosen to be unity.

\section{Abelian classical solutions}

\subsection{Iterative procedure}

Classical equations of motion are

\eq
\sum_{\mu} \im\tr\Bigl\{ \sigma^a \onabla_{\mu} U_{x\mu\nu} \Bigr\} =0~,
                  \label{classeq_sun}
\en

\noi where $U_{x\mu\nu}\in SU(2)$. For abelian solutions
$~U_{x\mu\nu}=\exp\Bigl( i\sigma_3\theta_{x\mu\nu}\Bigr)~$
eq.~(\ref{classeq_sun}) becomes

\eq
\sum_{\nu } \bpartial_{\nu}\sin \theta_{x\mu\nu} = 0~.
                  \label{classeq_u1}
\en

\noi Let us represent the plaquette angle $~\theta_{x\mu\nu}~$ in the 
form
\eq
\theta_{x\mu\nu} = {\ttheta}_{x\mu\nu} + 2\pi \cdot 
n_{x\mu\nu}; 
~~ -\pi < \ttheta_{x\mu\nu} \le \pi~, 
\en

\noi and $~n_{x\mu\nu}=-n_{x\nu\mu}~$ are integer numbers. The classical 
equations of motion (\ref{classeq_u1}) can be represented in the form

\eq
\sum_{\nu }\bpartial_{\nu} \theta_{x\mu\nu} =
F_{x\mu}(\theta)~,
                  \label{classeq1}
\en

\noi where
\eq
F_{x\mu}(\theta) \equiv \sum_{\nu }\bpartial_{\nu} 
\Bigl( \theta_{x\mu\nu}-\sin\theta_{x\mu\nu}\Bigr)
\nonumber 
\en

\noi and $~\sum_{\mu} \bpartial_{\mu} F_{x\mu}= 0~$.
For any given configuration $~\large\{ n_{x\mu\nu}\large\}~$
these equations can be solved iteratively 

\eq
\theta_{x\mu}^{(1)} \to \theta_{x\mu}^{(2)}\to 
\ldots \to \theta_{x\mu}^{(k)}\to \ldots~,
\en

\noi where
\eq
\sum_{\nu }\bpartial_{\nu} \theta_{x\mu\nu}^{(k+1)} =
F_{x\mu}(\theta^{(k)})~;  ~~k=1;~2;~\ldots~,
                \label{clas_1}
\en

\noi and

\eq
\sum_{\nu }\bpartial_{\nu} \theta_{x\mu\nu}^{(1)} =
2\pi\sum_{\nu }\bpartial_{\nu} n_{x\mu\nu}~;
                \label{clas_2}
\en

\noi In the Lorentz gauge 
$~\sum_{\mu} {\bar\partial}_{\mu} \theta_{x\mu} = 0~$
eq.'s~(\ref{clas_1},\ref{clas_2}) are equivalent to

\eq
\Delta \theta_{x\mu}^{(k+1)} = J_{x\mu}^{(k)}~;
~~k=0;~1;~\ldots~,
               \label{clas_3}
\en

\noi where $~J_{x\mu}^{(0)} = 2\pi\sum_{\nu }\bpartial_{\nu} 
n_{x\mu\nu}~$ and $~J_{x\mu}^{(k)} = F_{x\mu}(\theta^{(k)})~$ 
at $~k\ge 1~$. Evidently,

\eq
\sum_{\mu}\bpartial_{\mu}J_{x\mu}^{(k)} 
= 0~; \quad  \sum_x J_{x\mu}^{(k)} = 0~.
\en

\noi Defining the propagator

\eq
G_{x;y} = \frac{1}{V}\sum_{q\ne 0}
\frac{e^{iq(x-y)}}{{\cal K}^2}~;
~~ {\cal K}^2 = \sum_{\mu} 4\sin^2\frac{q_{\mu}}{2}~,
\en

\noi one can easily find solutions of eq.'s~(\ref{clas_3}) :

\eqa
\theta_{x\mu}^{(1)} &=& 2\pi \sum_{y\nu} G_{x;y} \cdot 
\bpartial_{\nu} n_{x\mu\nu}~;
           \label{first_iter}
\\
\nonumber \\
\theta_{x\mu}^{(k+1)} &=& \sum_y G_{x;y} \cdot J_{y\mu}^{(k)}~;
\quad \sum_{\mu} \bpartial_{\mu} \theta_{x\mu}^{(k+1)} = 0~,
\ena

\noi and $~\sum_x \theta_{x\mu}^{(k+1)} = 0$.

The results of iterative solution can be summarized as follows.

\begin{figure}[pt]
\begin{center}
\leavevmode
\hbox{
\epsfysize=14cm
\epsfxsize=14cm
\epsfbox{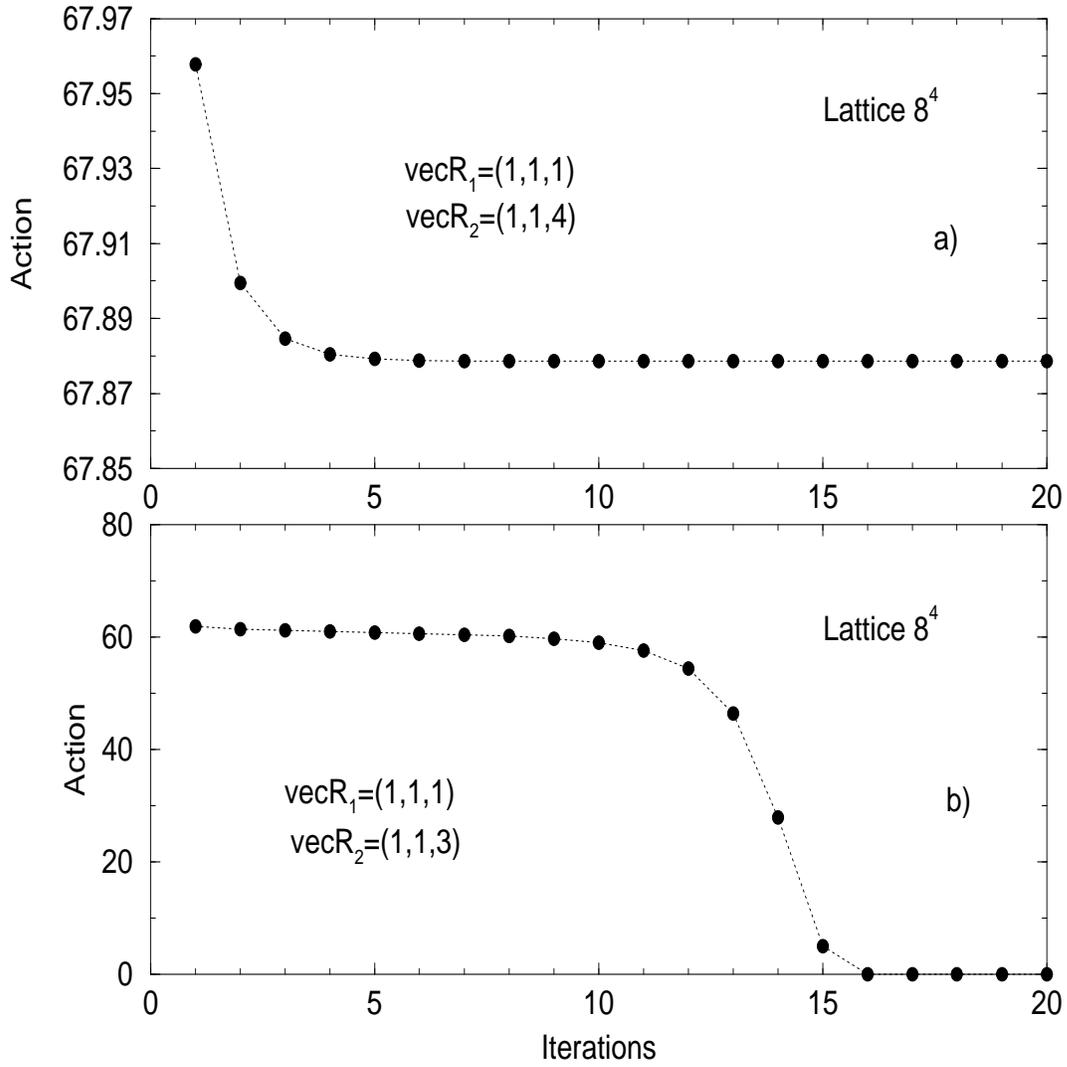}
}
\end{center}
\caption{Iterative solution of the classical equations}
\label{fig:iter}
\end{figure}

\begin{enumerate}

\item The convergence of this iterative procedure is very fast and 
becomes even faster with increasing the distance between monopole and 
antimonopole. As an example in Figure \ref{fig:iter}a it is shown the 
dependence of the action on the number of iterations on the $~8^4~$ 
lattice where ${\vec R}_1$ and ${\vec R}_2$ are positions of the static 
monopole and antimonopole, respectively. In fact, the first 
approximation $~\theta^{(1)}_{x\mu}~$ as given in eq.(\ref{first_iter})
is a very good approximation to the exact solution.

\item There are {\it no} solutions when monopole and antimonopole
are too close to each other. As an example, in Figure \ref{fig:iter}b
one can see the dependence of the action on the number of iteration
steps when ${\vec R}_2-{\vec R}_1=(0,0,2)$.

\item In four dimensions only {\it static} (i.e. 
threedimensional) solutions have been found.

\end{enumerate}

\subsection{Stability}

The question of stability of the classical solution $U^{cl}_{x\mu}$ is 
the question of the eigenvalues $\lambda_j$ of the matrix 
$~L^{ab}_{xy;\mu\nu}~$ where 

\eq 
S(e^{i\delta\theta}U^{cl}) = S_{cl} + \sum_{abxy\mu\nu} 
\delta\theta^a_{x\mu}L^{ab}_{xy;\mu\nu}\delta\theta^b_{y\nu}
+\ldots ,
\en

\noi where $S_{cl}=S(U^{cl})$ and $\delta\theta^a_{x\mu}$ are 
infinitesimal variations of the gauge field. A solution
$U^{cl}_{x\mu}$ is stable if all $\lambda_j\ge 0$. However,
a solution can be unstable but "quasistable" if, say, only one 
eigenvalue is negative : $\lambda_1 <0$, $~\lambda_j \ge 0, ~j\ne1$.
A cooling history of such configuration could have demonstrated
an approximate plateau.
If it could have been a case, one could extend, in principle, Polyakov's
approach to the case of quasistable solutions.

It is rather easy to show that in the case of $U(1)$ theory $M{\bar M}$ 
solutions are {\it stable}, i.e. correspond to local minima of the 
action.  Therefore, Polyakov's approach based on the $M{\bar M}$
classical solutions is expected to describe confinement and 
pseudoparticles are (anti)monopoles.

\begin{figure}[pt]
\begin{center}
\vskip -1.5truecm
\leavevmode
\hbox{
\epsfysize=14cm
\epsfxsize=14cm
\epsfbox{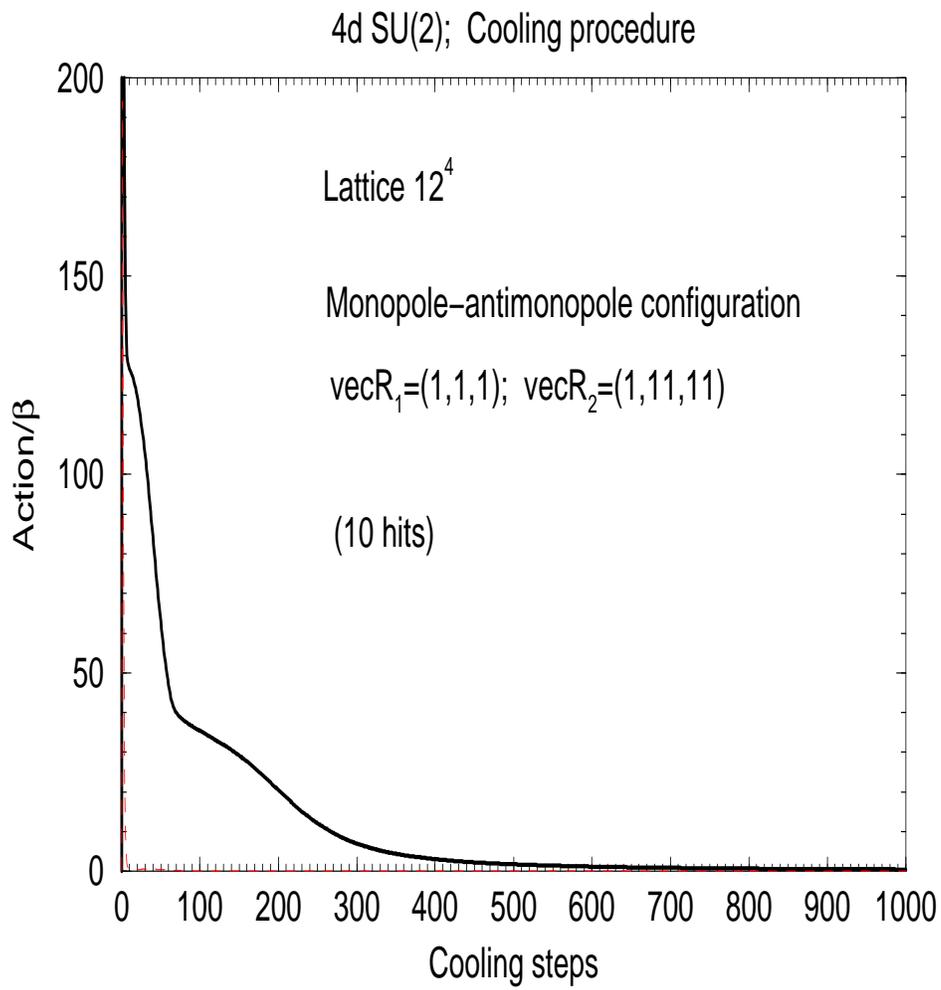}
}
\end{center}
\caption{Cooling history for the monopole-antimonopole pair}
\label{fig:cool_mmbar_4d}
\end{figure}

Stability of $M{\bar M}$--classical solutions in 
$SU(2)$ theory  has been studied numerically. To this purpose 
every classical $M{\bar M}$ configurations has been (slightly)
heated and then a (soft) cooling procedure has been used.
In Figure \ref{fig:cool_mmbar_4d} one can see a typical cooling
history of such configuration. The classical action $S_{cl}$
corresponding to the $M{\bar M}$ configuration is $\sim 130$.
Therefore, $M{\bar M}$--classical solution looks absolutely
{\it unstable}.

\begin{figure}[pt]
\begin{center}
\vskip -1.5truecm
\leavevmode
\hbox{
\epsfysize=14cm
\epsfxsize=14cm
\epsfbox{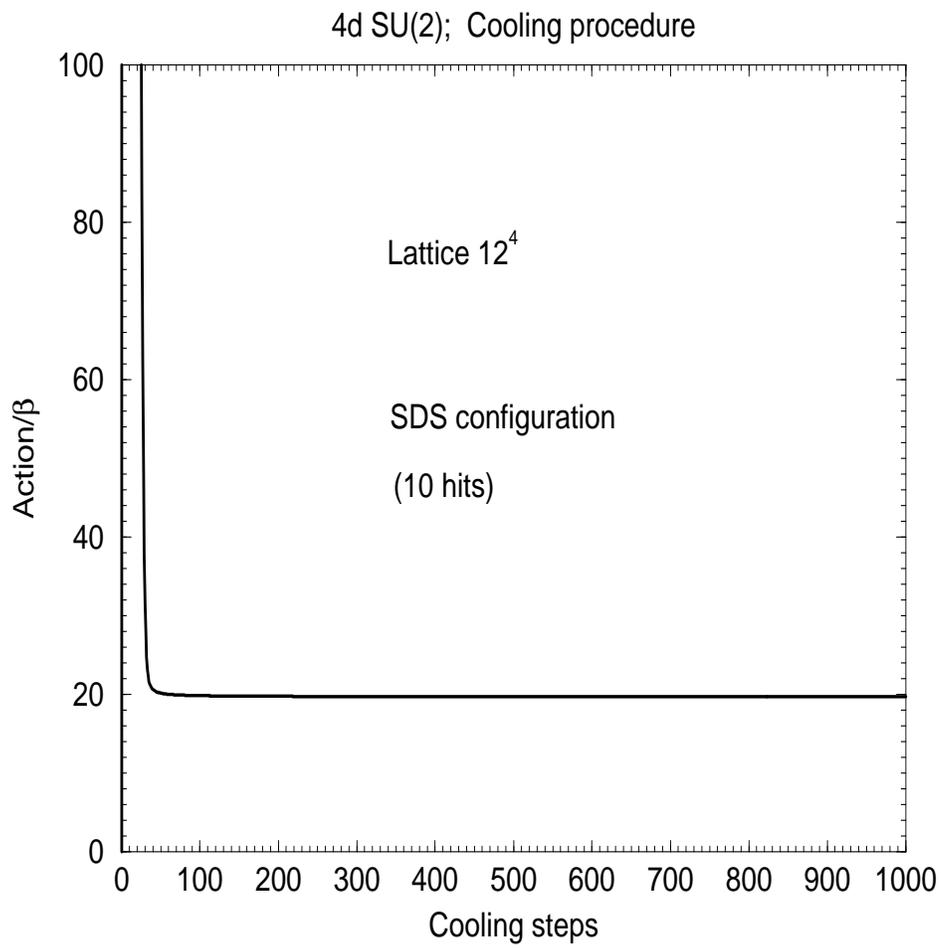}
}
\end{center}
\caption{Cooling history for the SDS}
\label{fig:cool_sds_4d}
\end{figure}

It is interesting to compare the stability of monopole
classical solutions with that of Dirac sheet (flux tube) solutions.
In Figure \ref{fig:cool_sds_4d} one can see a typical cooling
of the heated single Dirac sheet (SDS) \cite{ds} in $SU(2)$ theory.
Parameters of the cooling have been chosen the same for all
configurations. In fact, it is also unstable. However, a strong
plateau permits to define this configuration as a {\it quasistable}.

\section{Summary and discussions}

Classical solutions corresponding to monopole--antimonopole pairs in
$3d$ and $4d$ $SU(2)$ and (compact) $U(1)$ lattice gauge theories have
been found.

In the case of $~3d~$ and $~4d~$ $~U(1)~$ theories these
monopole--antimonopole classical solutions ($M{\bar M}$--pseudoparticles)
are {\it stable}, i.e. correspond to local minima of the action.
Therefore, the quasiclassical approach has  chance to be successful.

In contrast, in $~SU(2)~$ theory ($d=3$ and $d=4$) $M{\bar M}$ classical
solutions are completely unstable. At the moment it is not clear if
Polyakov's (quasiclassical) approach can be applied to nonabelian
theories (at least, with monopoles as pseudoparticles).  It is very
probable that the vacuum in the (compact) $U(1)$ theory is a rather poor
model of the vacuum in $~SU(2)~$ theory.

It is interesting to note that the Dirac sheet (i.e. flux tube)
solutions are quasistable in $~SU(2)~$ theories (for $d=3$ and $d=4$).
This observation could be interesting in view of the famous spaghetti
vacuum picture where the color magnetic quantum liquid state is a
superposition of flux--tubes states (Copenhagen vacuum) \cite{cv}.
However, the relevance of this scenario still needs a further confirmation.


%
%

\vfill


%
%
%

\vfill


%
%
%

\vfill

\end{document}